\documentclass[preprint,number]{elsarticle}

\begin{document}

\begin{frontmatter}



\title{Positive unlabeled learning for building recommender systems in a parliamentary setting}


\author[decsai]{Luis M. de Campos\corref{cor}}
\ead{lci@decsai.ugr.es}

\author[decsai]{Juan M. Fern\'andez-Luna}
\ead{jmfluna@decsai.ugr.es}

\author[decsai]{Juan F. Huete}
\ead{jhg@decsai.ugr.es}

\author[decsai]{Luis Redondo-Exp\'osito}
\ead{luisre@decsai.ugr.es}

\cortext[cor]{Corresponding author}

\address[decsai]{Departamento de Ciencias de la Computaci\'on e Inteligencia Artificial, \\ ETSI Inform\'atica y de Telecomunicaci\'on, CITIC-UGR, \\ Universidad de Granada, 18071, Granada, Spain}

\begin{abstract}

Our goal is to learn about the political interests and preferences of the Members of Parliament by mining their parliamentary activity, in order to develop a recommendation/filtering system that, given a stream of documents to be distributed among them, is able to decide which documents should receive each Member of Parliament.
We propose to use positive unlabeled learning to tackle this problem, because we only have information about relevant documents (the own interventions of each Member of Parliament in the debates) but not about irrelevant documents, so that we cannot use standard binary classifiers trained with positive and negative examples. We have also developed a new algorithm of this type, which compares favourably with: a) the baseline approach assuming that all the interventions of other Members of Parliament are irrelevant, b) another well-known positive unlabeled learning method and c) an approach based on information retrieval methods that matches documents and legislators' representations. The experiments have been carried out with data from the regional Andalusian Parliament at Spain.
\end{abstract}


\end{frontmatter}


\section{Introduction}
\label{intro}

Nowadays we live in the information society, where enterprises, institutions and people in general can easily access to vast amounts of information. Moreover, in many cases users do not need to actively search for the information they need but they play a more passive role and are constantly bombarded with advertising, news, e-mails, etc. The problem is then to separate the chaff from the wheat, to distinguish what is interesting, important or useful and what is not. This is a hard and time-consuming task. To reduce the information overload, there exist content-based recommender/filtering systems \cite{belkin92,pazzani07}, which suggest items (songs, movies, books, restaurants,...) to users according to their preferences and taking into account the characteristics of the items.

This situation also happens in a political context. Politicians in general and Members of Parliament (MP) in particular need to keep informed about the matters more related with their specific political interests. For example, an MP who is working in the health committee of a regional or national parliament probably would be interested in documents produced by the European Union dealing with health-related matters but not in others concerning, let us say, education or agriculture. In our case the users will be the MPs and the items to be recommended/filtered are the documents arriving to the parliament (e.g. news releases or technical reports). The goal is to develop a system which can automatically decide those MPs who should receive each document. This decision must be based on both the content of the document and the political interests of each MP.

One possible approach to develop our recommendation/filtering system could be to learn about the interests and preferences of each MP by mining her parliamentary activity. So, we could use the transcriptions of the speeches of the MPs in the parliamentary debates to train a binary classifier (the class values being relevant and non-relevant) for each MP. Then, when a new document to be filtered/recommended enters the system, we could use these classifiers to determine which MPs should receive the document: those MPs whose associated classifier predicts the relevant class. Alternatively, if we assume that the classifiers produce a numerical output instead of a binary value, we could generate a ranking of MPs in decreasing order, thus recommending the document to the top ranked MPs.

The problem with this approach is that in order to build a standard binary classifier for each MP we need training data (documents in this case), both positive (relevant documents) and negative (irrelevant documents). Positive training data do not represent any problem: the own interventions/speeches of an MP are clearly positive training data for building the classifier for this MP. However, negative training data are not so clear. We could suppose that all the interventions which are not from an MP are negative training data for the classifier associated to this MP. But this may be somewhat unreasonable, because the interventions of other MPs dealing with the topics which are of interest for a given MP could also be relevant for this MP, thus creating confusion in the classifier. For example, if a given MP is interested or especialized in education, it is quite probable that she will find relevant (at least some of) the interventions of other MPs about this topic. Therefore, within the interventions of other MPs we will probably find both relevant and irrelevant documents for a given MP.

This situation can be managed using the techniques known as Positive Unlabeled Learning (PUL) \cite{zhang08}, where it is assumed that there exists a set of positive data and a (usually larger) set of unlabeled data, but there is no negative training data. In our case the unlabeled data would correspond with the interventions of all the other MPs. PUL is an extreme case of semisupervised learning \cite{SSL06} (which considers simultaneously positive, negative and unlabeled data).

So, our proposal in this paper is to explore the use of positive unlabeled learning to build a content-based recommender system of documents for the MPs. More precisely, our approach is first based on trying to detect, among the unlabeled data, a subset of reliable negative data, and second to use the known positive data and the reliable negative data to train a standard binary classifier for each MP. To detect reliable negative data we can use some of the known PUL methods, although we propose a new method based on constraining the operations of the K-means clustering algorithm.

In order to validate our proposals, we shall perform an experimental study using a collection of MPs interventions from the regional Parliament of Andalusia at Spain.

The main contributions of this paper are: (a) the proposal of using machine learning techniques to tackle the problem of building a content-based recommender system of documents in a parliamentary setting (there are other proposals to deal with this problem \cite{egovis15,kdir15,JIS16}, but all using information retrieval-based methods instead of machine learning techniques); (b) the use of positive unlabeled learning to build a recommender system (we are not aware of any work in this sense, although there are many papers applying positive unlabeled learning to the problem of classifying documents \cite{denis02,fung06,li03,liu02,ICDM03,yu02}); (c) the proposal of a new method of positive unlabeled learning based on a modification of the K-means clustering algorithm.

The rest of the paper is organized in the following way: in Section \ref{related} we summarize related work. Section \ref{main} describes in detail our approach. Section \ref{experiments} contains the experimental part of the paper. Finally, Section \ref{concluding} contains the concluding remarks and some proposals for future work.

\section{Related work}
\label{related}

There are many works studying the recommendation/filtering problem in different domains and applications (as the three survey papers \cite{hanani01,surveyRS1,surveyRS2} show). Content-based recommender systems can be built using either information retrieval-based methods \cite{baeza,belkin92,foltz92,loeb92,narducci16} or machine learning algorithms for learning user models \cite{billsus02,cohen96,kim01,jennings93,pazzani97,tjoa97,zahra15}. However, its application in a parliamentary context is much more limited \cite{egovis15,kdir15,JIS16}, and in all these cases only information retrieval-based methods have been used.

In \cite{egovis15} a lazy approach is considered, avoiding to construct an elaborated profile of the MPs, collecting the transcriptions of all their speeches and building a document collection, then using an information retrieval system to search the MPs most similar to the document to be recommended. This approach is refined in \cite{JIS16}, where term (word) profiles for the different MPs are extracted from their speeches in different ways. A different approach is considered in \cite{kdir15}, where the profiles of the MPs are not built from the terms in their interventions but from the keywords manually assigned by documentalists (using a thesaurus) to these interventions.

On the other hand, there are three classes of methods proposed for positive unlabeled learning, according to \cite{zhang08}. The first class uses a two-step strategy, where the first step tries to identify a set of reliable negative data from the unlabeled set, and the second step uses a traditional supervised learning algorithm on the positive and the reliable negative data \cite{li03,liu02,ICDM03,yu02}. The second class follows the statistical query learning model. For example, in \cite{denis02} a modification of the Naive Bayes (NB) for text classification is obtained by estimating the conditional probabilities of the terms given the positive class in the usual way and the conditional probabilities given the negative class by using a supplied estimate of the prior probability of the positive class. In \cite{calvo07} other Bayesian network classifiers are also extended to the PUL setting. The third class of methods treats the unlabeled data as noisy negative examples, using then logistic regression \cite{lee03} or the Biased Support Vector Machine \cite{ICDM03}, for example. PUL is also being used in the case of data streams \cite{liang12}, and it is still an active area of research \cite{plessis17,gan17,hernandez17}.

We are going to focus on the methods of the first class, which are more extended and are more similar to the new PUL method that we propose. In \cite{ICDM03} the authors use the NB classifier, where positive data are used as positive training examples and unlabeled data as negative training examples. The resulting NB classifier is used to re-classify the unlabeled data, thus selecting as reliable negative data those unlabeled examples which have been classified as negative by NB. A similar approach is used in \cite{li03}, where NB is replaced by the Rocchio text classification method (using tf-idf weights and the cosine similarity). Another proposal is the Spy technique \cite{liu02}, which randomly selects a subset of positive data to be added to the unlabeled data. Then the Expectation Maximization (EM) algorithm is applied to train a NB classifier, which is used on the selected positive data to obtain a threshold able to identify reliable negative examples. The PEBL method \cite{yu02} tries to identify those features (terms in this case), called positive features, which are more frequent (in relative terms) between positive documents than between unlabeled documents. Then those documents which do not contain any of these positive features are selected as reliable negative examples. There are also proposals (e.g. \cite{fung06}) that try to obtain from the unlabeled data both reliable positive and negative data.

\section{Positive unlabeled learning in a parliamentary setting}
\label{main}

The situation that we are considering can be formalized as follows: let ${\cal MP} = \{MP_1,\ldots,MP_n\}$ be a set containing all the MPs working in a parliament. This institution receives or generates a series of documents that should be distributed among the MPs. However, to alleviate their  work, not all the MPs should receive all the documents \cite{shamin07}. Instead, each MP should receive only those documents which are related to her interests, preferences and the role she plays within the parliament. Therefore, a system able to automatically perform this filtering process is required. As mentioned in Section \ref{intro}, we want to build such a system by using machine learning techniques, more precisely positive unlabeled learning. The (public and, we expect, reliable) source of information about the political interests of MPs will be their interventions within the parliamentary debates. So, each MP$_i$ can be associated with a set of documents ${\cal D}_i = \{d_{i1},\ldots,d_{im_i}\}$, where each $d_{ij}$ represents the transcription of the speech of MP$_i$ when she participated in the discussion of a parliamentary initiative. The complete set of documents is ${\cal D} = \cup_{j=1}^n{\cal D}_j$. Therefore, we are going to train a set of $n$ binary text classifiers using ${\cal D}$. For each MP$_i$, the set of positive examples (documents) is precisely ${\cal D}_i$, whereas the set of unlabeled documents is ${\cal D}\setminus {\cal D}_i$.

Our proposal for using PUL to build a recommender/filtering system of documents for MPs falls within the two-step strategy mentioned in Section \ref{related}. We shall use a modification of the K-means clustering algorithm in the first step, in order to identify a set of reliable negative documents, ${\cal N}_i$, from the set of unlabeled documents for each MP$_i$ (i.e. the interventions of other MPs).

In the second step, for each MP$_i$ we will train a binary classifier from ${\cal D}_i$ and ${\cal N}_i$ using Support Vector Machines \cite{SVM}, which is considered as the state-of-the-art technique for document classification. As it is quite probable that the sets ${\cal N}_i$ are quite larger than the corresponding ${\cal D}_i$, i.e. the data sets can be quite imbalanced, we have also considered the possibility of using some method to deal with the class imbalance problem.

\subsection{The modified K-means algorithm}

The classical K-means algorithm is an iterative method that, starting from an initial centroid for each of the K clusters, assigns each example to the cluster whose centroid is nearer (more similar) to the example. Then the algorithm recomputes the centroid of each cluster using all the examples assigned to it. The new centroids are used to reassign each example to the (possibly different) cluster whose centroid is more similar to the example, and this process is repeated until a convergence condition holds. In our case the number of clusters is fixed to K=2 and the similarity between documents is computed using the classical cosine similarity measure \cite{baeza}. The proposed modification is that the known positive examples are forced to always remain in the positive cluster, no matter if they are more similar to the negative centroid, whereas the unlabeled examples can fluctuate between the two clusters depending on the similarity. To initialize the process, the positive centroid is computed from all the positive examples and the negative centroid is calculated from all the unlabeled examples. At the end of the process, the unlabeled examples which remain in the negative cluster are considered reliable negative examples.

\section{Experimental evaluation}
\label{experiments}

To experimentally evaluate our proposals, we shall use data from the Andalusian Parliament at Spain\footnote{http://www.parlamentodeandalucia.es}. More precisely, we focus on the 8th term of office of this regional chamber, where a total of 5,258 parliamentary initiatives were discussed.

Each initiative, marked up in XML \cite{campos09}, includes the transcriptions of all the speeches of the MPs who participate in the discussion, together with their names. There is a total of 12,633 different interventions (with an average of 2.4 interventions per initiative). Our set ${\cal MP}$ is composed of 132 MPs\footnote{We considered only those MPs who intervene in at least 10 initiatives.}.

We randomly partitioned the set of initiatives into a training set (containing 80\% of the initiatives) and a test set (containing the remaining 20\%). To obtain more statistically reliable results, we repeated this process 5 times, and the reported results are the averages of these rounds. In other words, we used the repeated holdout method \cite{lantz13} as evaluation methodology.

We extracted the interventions of all the MPs in ${\cal MP}$ from the initiatives in the training set and used them to build a classifier for each MP, according to the method described in Section \ref{main}. These classifiers were then used to classify the initiatives in the test set, using the transcriptions of the speeches within each test initiative as the document to be filtered/recommended, assuming that each test initiative is relevant only for those MPs who participate in it. It is worth to mention that this is a very conservative assumption, because an initiative could also be relevant to other MPs who did not participate in its discussion but are interested in the same topics the initiative is devoted to. Our assumption is an easy way to establish a kind of ``ground truth'', without the need to have documents annotated with explicit relevance judgements.

In order to assess the quality of the filtering/recommendation system we used classical evaluation measures of text classification, namely precision, recall and the F-measure \cite{sebastiani02}. Let $TP_i$ (True Positives) be the number of test initiatives which are truly relevant for MP$_i$ and have been classified as relevant by the classifier associated to MP$_i$; $FP_i$ (False Positives) is the number of test initiatives which are not relevant for MP$_i$ but have been incorrectly identified as relevant by the corresponding classifier; $FN_i$ (False Negatives) is the number of test initiatives that, although being relevant for MP$_i$, have been incorrectly classified as irrelevant. Precision is then defined as $p_i=TP_i/(TP_i+FP_i)$ (an estimation of the probability of a document being truly relevant given that it is classified as relevant. Recall is defined as $r_i=TP_i/(TP_i+FN_i)$ (an estimation of the probability of classifying as relevant a truly relevant document). The F-measure is the harmonic mean of precision and recall, $F_i=2p_ir_i/(p_i+r_i)$.

As we compute precision, recall and F for every MP$_i$, it is necessary to summarize each of these three types of measures into a single value which gives a global perspective of the system's performance. To this end we used both macro-averaged (Mp, Mr and MF) and micro-averaged (mp, mr and mF) measures \cite{Tsoumakas10}:

\begin{equation}
Mp = \frac{1}{n}\sum_{i=1}^n p_i, \qquad Mr = \frac{1}{n}\sum_{i=1}^n r_i, \qquad MF = \frac{1}{n}\sum_{i=1}^n F_i
\label{macro}
\end{equation}

\begin{equation}
mp = \frac{\sum_{i=1}^n TP_i}{\sum_{i=1}^n (TP_i+FP_i)}, \;\; mr = \frac{\sum_{i=1}^n TP_i}{\sum_{i=1}^n (TP_i+FN_i)}, \;\; mF = \frac{2mp\,mr}{mp+mr}
\label{micro}
\end{equation}

The baseline approach ({\it bas}) we have considered is to train the classifiers without using PUL, i.e. for each MP$_i$ the set of positive examples is again ${\cal D}_i$, whereas all the unlabeled examples in ${\cal D}\setminus {\cal D}_i$ are considered as negative examples. We shall also use, for comparison purposes, the well-known PUL method proposed in \cite{ICDM03} (and described in Section \ref{related}), based on Naive Bayes ({\it pul-nb}). Once the reliable negative examples are identified by this method, SVMs are also used to build the classifiers. The method based on using the modification of K-means, proposed in Section \ref{main} as the first step of PUL will be called {\it pul-km}. The comparison of {\it bas} and {\it pul-km} will serve to assess the merits of PUL in our recommendation context. The comparison of {\it pul-km} and {\it pul-nb} will give us an idea of the potential of the new PUL method proposed in this paper.

As mentioned in Section \ref{main}, we shall also experiment with versions of {\it bas}, {\it pul-nb} and {\it pul-km} (called {\it bas-b}, {\it pul-nb-b} and {\it pul-km-b} respectively) where, previous to applying SVMs to the sets of positive and (reliable) negative examples, we use a method to deal with the imbalance of these data sets. More precisely, we have used the synthetic minority over-sampling technique (SMOTE) \cite{smote}, which essentially is a statistical algorithm for creating new instances, from existing cases of the minority class. SMOTE works by taking samples of the class with less observations and its k nearest neighbors randomly. Then it produces new observations setting a random point along the segment generated between the target sample and its k neighbors. We used the implementations of SVM, NB and SMOTE available in R\footnote{https://cran.r-project.org} (packages {\it caret}, {\it e1071} and {\it DMwR}). All the preprocessing steps of the datasets (all the initiatives were preprocessed by removing stop words and performing stemming) were also carried out with R packages ({\it tm} and {\it snowBallC}). The modified K-means algorithm and the evaluation process were implemented in Java.

The version used of the selected classification algorithm (SVM) is able to give a numerical output, more precisely it returns the probability of the target document $d$ being relevant to MP$_i$, $pr_i(d)$. Thus, we can use it by simply assigning the relevant value to $d$ if $pr_i(d)\geq 1-pr_i(d)$ (i.e. if $pr_i(d)\geq 0.5$). But, more generally, we can also select a threshold $t$ ($0\leq t\leq 1$) and state that $d$ is relevant for MP$_i$ if $pr_i(d)\geq t$. In this sense the values of $TP_i$, $FP_i$ and $FN_i$ used to compute precision and recall are obtained according to the contingency table displayed in Table \ref{table-contingency}. We have experimented with several values for the threshold $t$, ranging from 0.1 to 0.9.

\begin{table}[htb]
\begin{center}
\begin{tabular}{c|cc}
                      & Truly relevant & Truly irrelevant \\
                      & for MP$_i$     & for MP$_i$ \\  \hline
$pr_i(d)\geq t$ & $TP_i$         & $FP_i$ \\
$pr_i(d)< t$   & $FN_i$         &  \\ \hline
\end{tabular}
\end{center}
\caption{Contingency table for threshold $t$.}
\label{table-contingency}
\end{table}

\subsection{Results with imbalanced data sets}

\begin{figure}[htb]
\begin{center}
\includegraphics[width=0.65\textwidth]{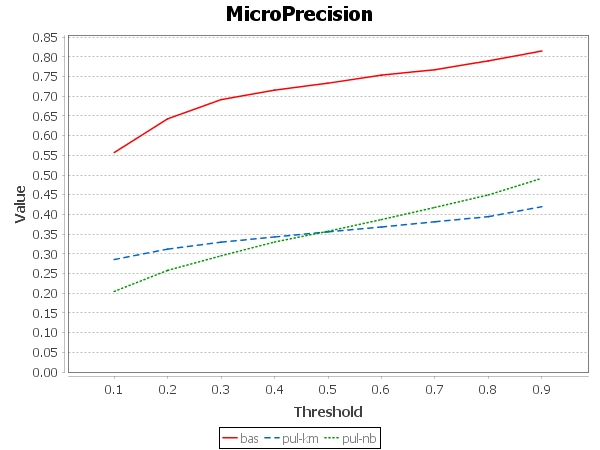}

\vspace{0.1cm}
\includegraphics[width=0.65\textwidth]{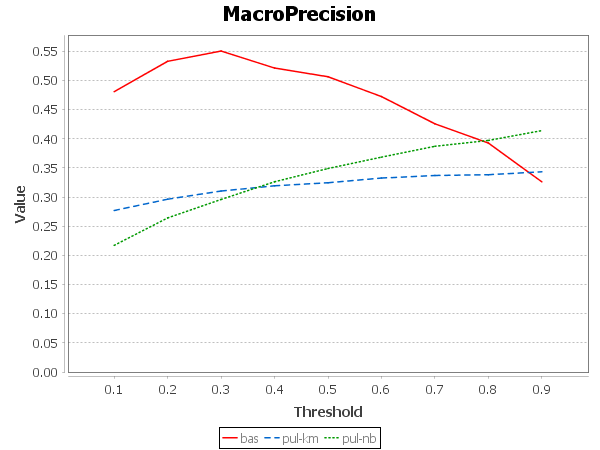}
\caption{Micro and Macro precision for {\it bas}, {\it pul-km} and {\it pul-nb} using different thresholds.} 
\label{fig1}
\end{center}
\end{figure}

\begin{figure}[htb]
\begin{center}
\includegraphics[width=0.65\textwidth]{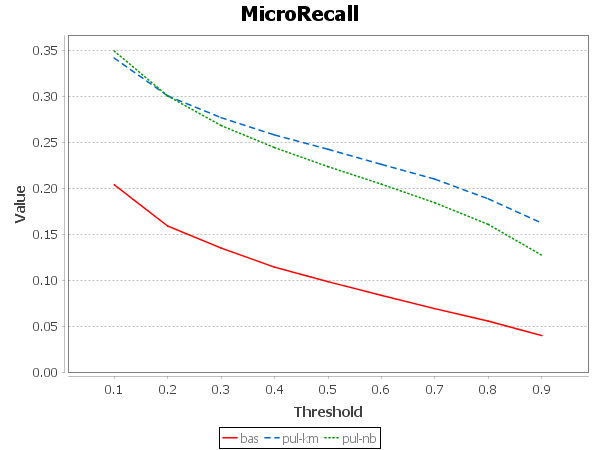}

\vspace{0.1cm}
\includegraphics[width=0.65\textwidth]{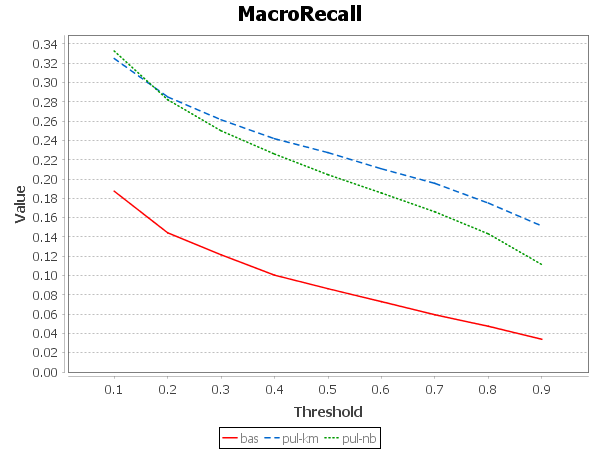}
\caption{Micro and Macro recall for {\it bas}, {\it pul-km} and {\it pul-nb} using different thresholds.} 
\label{fig2}
\end{center}
\end{figure}

\begin{figure}[htb]
\begin{center}
\includegraphics[width=0.65\textwidth]{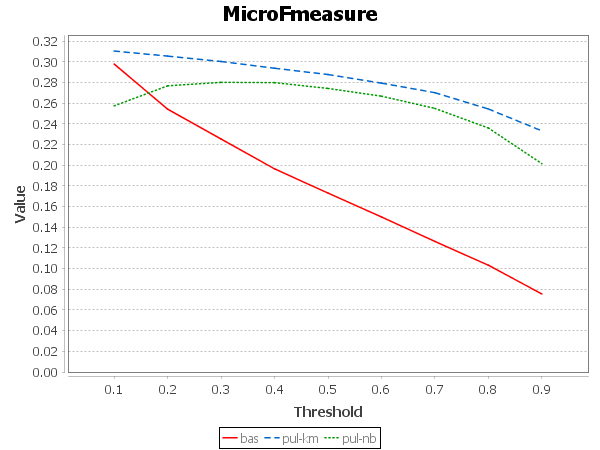}

\vspace{0.1cm}
\includegraphics[width=0.65\textwidth]{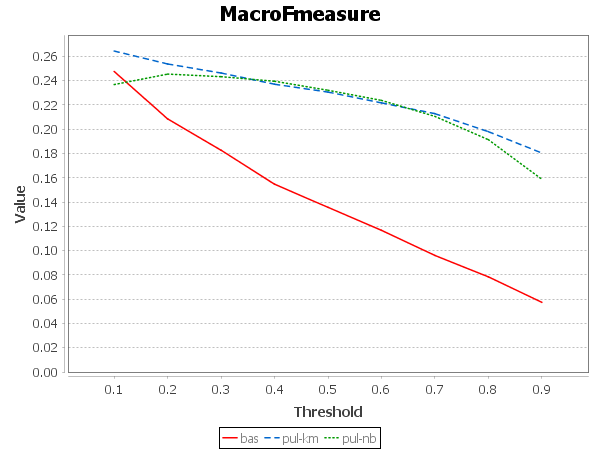}
\caption{Micro and Macro F measures for {\it bas}, {\it pul-km} and {\it pul-nb} using different thresholds.} 
\label{fig3}
\end{center}
\end{figure}

The results of our experiments for (micro and macro) precision, recall and F using different thresholds are displayed in Figures \ref{fig1} to \ref{fig3}, respectively.

First, the results in Figures \ref{fig1} and \ref{fig2} allow us to extract some general tendencies for the three approaches: precision increases and recall decreases as the threshold increases. This is to be expected. When the threshold increases the classifiers are more selective to assign the relevant value to a document. The consequence is that the number of false positives decreases, thus increasing precision. At the same time, the number of false negatives increases and therefore recall decreases. The exception is the behaviour of macro precision with the {\it bas} approach: this measure tends to decrease when increasing the threshold. We believe that this reveals a poor performance of this approach in those cases where the classifiers are trained with very few positive examples, i.e. for those MPs that scarcely intervene in the debates (then having a heavily imbalanced training set). In these cases, although the number of false positives decreases as the threshold increases, the number of true positives also decreases more quickly. Notice that this only affects macro precision and not micro precision, because in the first case all the MPs contribute equally to this measure, independently on their number of interventions.

These figures also show that the baseline approach and the two PUL methods behave differently: the {\it bas} approach is much better for precision and the PUL methods are much better for recall. Given the characteristics of our evaluation method, we believe that we should give more importance to recall than to precision. The reason is that false negatives (which affect recall) represent true errors, an MP participated in an initiative but the classifier does not recommend this to her. On the other hand, a false positive (which affects precision) represents that an MP did not participate in an initiative but the classifier recommends this to her. It could be the case that this MP is truly interested in this initiative because its content matches with her political interests. In this way, low recall is an objective signal of bad performance, whereas low precision does not necessarily means the same, it may be a by-product of our conservative assumption concerning relevance.

In Figure \ref{fig3} we can observe the results for the F-measure (micro and macro), which represents a balance between precision and recall, and therefore is an appropriate measure of global performance. First, we can see that the best results are always obtained when we use low thresholds. Second, {\it pul-km} systematically outperforms both {\it bas} and {\it pul-nb}.

Table \ref{table-imbalanced} shows the best F values obtained by each approach, as well as the corresponding thresholds where these values are reached. We have used paired t-tests (using the results of the five random partitions, and a confidence level of 95\%) to assess the statistical significance of these results. {\it pul-km} is always significantly better than both {\it bas} and {\it pul-nb}. At micro level {\it bas} is also significantly better than {\it pul-nb}, whereas there is not significant difference between these two approaches at macro level.

\begin{table}[htb]
\begin{center}
\begin{tabular}{|c|ccc|} \hline
Approach & bas & pul-km & pul-nb \\ \hline
 &   & Micro-F & \\
Value    & 0.2978 & {\bf 0.3105} & 0.2802 \\
Threshold & 0.1 & 0.1 & 0.3 \\ \hline
 &   & Macro-F & \\
Value & 0.2475 & {\bf 0.2644} & 0.2454 \\
Threshold & 0.1 & 0.1 & 0.2 \\ \hline
\end{tabular}
\end{center}
\caption{Best micro and macro F values obtained by {\it bas}, {\it pul-km} and {\it pul-nb}.}
\label{table-imbalanced}
\end{table}

\subsection{Results with balanced data sets}

\begin{figure}[htb]
\begin{center}
\includegraphics[width=0.65\textwidth]{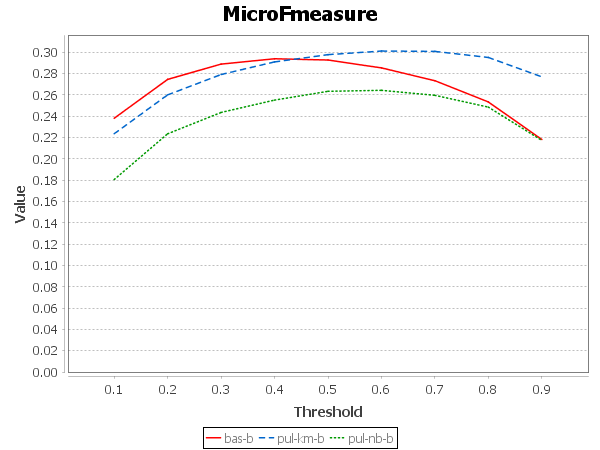}

\vspace{0.1cm}
\includegraphics[width=0.65\textwidth]{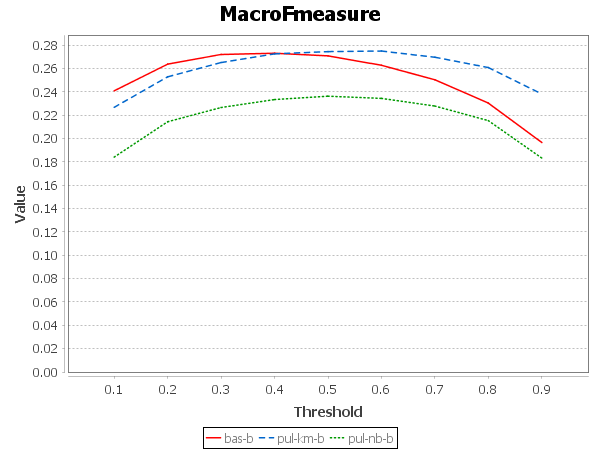}
\caption{Micro and Macro F measures for {\it bas-b}, {\it pul-km-b} and {\it pul-nb-b} using different thresholds.} 
\label{fig-balF}
\end{center}
\end{figure}

Now, we are going to repeat the experiments of the previous section but using the balanced versions of the three approaches, {\it bas-b}, {\it pul-km-b} and {\it pul-nb-b}. For the sake of conciseness, we only show the results relative to the F-measure in Figure \ref{fig-balF}. The figures for precision and recall exhibit a behaviour similar to those in the previous section, increasing lines for precision and decreasing lines for recall, although the lines are closer. Also, the previous extrange behaviour of macro precision with the {\it bas} approach has dissappeared.

In Figure \ref{fig-balF} we can observe several interesting facts. First, the thresholds where the best results are obtained have changed completely, now they are more centered, near the point 0.5 which could be considered as the natural threshold. This seems to indicate that the classifiers are better calibrated, they do not need to draw on very low thresholds to get good results. Second, {\it pul-km-b} is still the best approach, although the differences with {\it bas-b} are smaller than in the previous experiments. Third, balancing {\it pul-nb} is not a good idea, it obtains results considerably worse than {\it pul-km-b} and {\it bas-b}. Table \ref{table-balanced} is the counterpart of Table \ref{table-imbalanced} for the balanced case. Comparing Figures \ref{fig-balF} and \ref{fig3} and Tables \ref{table-balanced} and \ref{table-imbalanced} we can see that balancing the data sets improves macro F (except for {\it pul-nb-b}) but systematically deteriorates the best values of micro F. We believe that this behaviour is caused because balancing improves the classifiers associated to MPs having a low number of interventions, but may worsen those MPs with a greater number of interventions (which are those more influencing in the value of micro F). The t-tests in this case indicate that there are not significant differences between {\it pul-km-b} and {\it bas-b}, but both {\it pul-km-b} and {\it bas-b} are significantly better than {\it pul-nb-b}.

\begin{table}[htb]
\begin{center}
\begin{tabular}{|c|ccc|} \hline
Approach & bas-b & pul-km-b & pul-nb-b \\ \hline
 &   & Micro-F & \\
Value    & 0.2940 & {\bf 0.3012} & 0.2643 \\
Threshold & 0.4 & 0.6 & 0.6 \\ \hline
 &   & Macro-F & \\
Value & 0.2732 & {\bf 0.2751} & 0.2364 \\
Threshold & 0.4 & 0.6 & 0.5 \\ \hline
\end{tabular}
\end{center}
\caption{Best micro and macro F values obtained by {\it bas-b}, {\it pul-km-b} and {\it pul-nb-b}.}
\label{table-balanced}
\end{table}

\subsection{Results when increasing the number of initiatives where MPs must intervene}

In all the previous experiments, we have built classifiers for all the MPs who participate in at least 10 initiatives. This constitutes a very heterogeneous set of MPs: there are MPs which participate in hundreds of initiatives and other, more passive, that scarcely intervene in the debates. Our goal in this section is to evaluate the proposed approaches when we impose a greater limit to the number of initiatives where MPs must intervene in order to be included in the study.

\begin{figure}[htb]
\begin{center}
\includegraphics[width=0.65\textwidth]{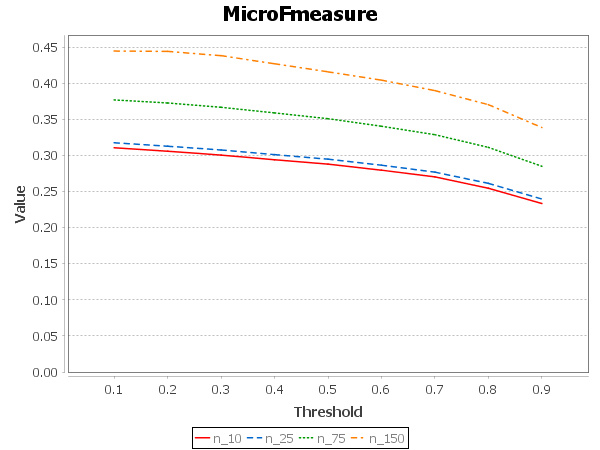}

\vspace{0.1cm}
\includegraphics[width=0.65\textwidth]{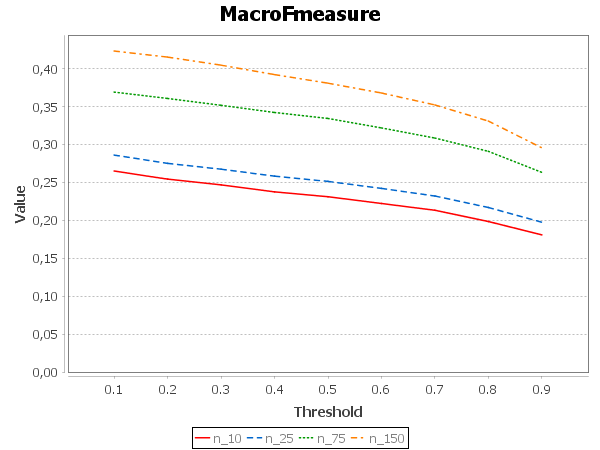}
\caption{Micro and Macro F measures for {\it pul-km} using different thresholds, for a minimum of 10, 25, 75 and 150 interventions.} 
\label{fig-F-interv}
\end{center}
\end{figure}

Therefore, we have repeated the experiments but including only those MPs who participate in at least 25, 75, and 150 initiatives. Our hypothesis is that the results in these cases will be progressively better, because a greater number of interventions will exclude those MPs whose classifiers are less accurate due to the use of poor training sets. Table \ref{table-all} displays the best F values for the three approaches (in both the imbalanced and balanced cases). We also show in Figure \ref{fig-F-interv}, the micro and macro F measures obtained by {\it pul-km} using different thresholds\footnote{We do not show the corresponding figures for the other approaches to save space, but the trends are completely similar.}.

\begin{table}[htb]
\begin{center}
\begin{tabular}{|c|cccccc|} \hline
Approach & bas & bas-b & pul-km & pul-km-b & pul-nb & pul-nb-b\\ \hline
 & & & Micro-F & & & \\
mF10 & 0.2978 & 0.2940 & {\bf 0.3105} & 0.3012 & 0.2802 & 0.2643 \\
mF25 & 0.3037 & 0.3038 & {\bf 0.3175} & 0.3084 & 0.2859 & 0.2705 \\
mF75 & 0.3558 & 0.3597 & {\bf 0.3768} & 0.3647 & 0.3437 & 0.3072 \\
mF150 &0.4408 & 0.3987 & {\bf 0.4446} & 0.4171 & 0.4183 & 0.3584 \\ \hline
 & & & Macro-F & & & \\
MF10 & 0.2475 & 0.2732 & 0.2644 & {\bf 0.2751} & 0.2454 & 0.2364 \\
MF25 & 0.2658 & 0.2920 & 0.2863 & {\bf 0.2941} & 0.2630 & 0.2511 \\
MF75 & 0.3355 & 0.3563 & {\bf 0.3694} & 0.3629 & 0.3361 & 0.2887 \\
MF150 &0.4039 & 0.3761 & {\bf 0.4236} & 0.3984 & 0.3976 & 0.3407 \\ \hline
\end{tabular}
\end{center}
\caption{Best micro and macro F values obtained by {\it bas}, {\it bas-b}, {\it pul-km}, {\it pul-km-b}, {\it pul-nb} and {\it pul-nb-b} with different minimum number of interventions.}
\label{table-all}
\end{table}

We can see in Table \ref{table-all} that indeed the results with all the approaches systematically improve as the number of interventions required increases (this fact is also confirmed in Figure \ref{fig-F-interv}). We can also observe that the relative merits of each approach remain unmodified: {\it pul-km} is the best approach followed by {\it bas}, being {\it pul-nb} the worst. It is also apparent that balancing the data sets is always counterproductive for the micro F measure. Moreover, when the MPs being considered have a great number of interventions (75 and 150) balancing is not useful either for the macro F measure.

\subsection{Comparison with information retrieval-based approaches}

In this section we are going to compare our proposed approach, {\it pul-km}, with two of the information retrieval-based approaches mentioned in Section \ref{related} \cite{egovis15}. They use the documents in ${\cal D}$ to feed an Information Retrieval System (IRS)\footnote{In our experiments we have used the implementation in the search engine library Lucene (https://lucene.apache.org) of the BM25 information retrieval model.}. In both cases the document to be filtered/recommended is used as a query to the IRS, which returns a ranked list of the MPs which are more similar to it.

In one case the documents to be indexed by the IRS are all the interventions of all the MPs in the training set, i.e. just the documents in ${\cal D}$. We call this approach {\it ir-i}. In the other case we first build a kind of profile for each MP, by grouping together all her interventions in a single document (all the documents in ${\cal D}_i$ form a single document $d_i=\cup_{j=1}^{m_i}d_{ij}$). Then these ``macro'' documents are indexed by the IRS. We call this second approach {\it ir-p}. In both cases, as each document is unambiguously associated with an MP, we can replace the ranking of documents by a ranking of MPs. However, in the {\it ir-i} approach the ranking of MPs may contain duplicate MPs having different scores (which correspond to different interventions of the same MP). Therefore, in this case we remove all the occurrences of an MP except the one having the maximum score.

Notice that the scores returned by the IRS are affected by the number of terms in the query. As we are using a single threshold to recommend a document to those MPs whose score is greater than the threshold, we need to normalize the scores by dividing by the maximum score. In this way we make the range of the scores independent on the query.

Table \ref{table-ir} displays the best F values for the the two IR-based approaches (we repeat in the table the results for {\it pul-km} to ease the comparison).

\begin{table}[htb]
\begin{center}
\begin{tabular}{|c|ccc|ccc|} \hline
 & & Micro-F & & & Macro-F & \\ \hline
Approach & pul-km & ir-i &ir-p & pul-km & ir-i &ir-p \\ \hline
10 & {\bf 0.3105} & 0.2896 & 0.2892 & {\bf 0.2644} & 0.2423 & 0.2513 \\
25 & {\bf 0.3175} & 0.2971 & 0.2939 & {\bf 0.2863} & 0.2661 & 0.2829 \\
75 & {\bf 0.3768} & 0.3509 & 0.3085 & {\bf 0.3694} & 0.3288 & 0.3368 \\
150 & {\bf 0.4446}& 0.4282 & 0.3120 & {\bf 0.4236} & 0.3948 & 0.3530 \\ \hline
\end{tabular}
\end{center}
\caption{Best micro and macro F values obtained by {\it pul-km}, {\it ir-i} and {\it ir-p}, with different minimum number of interventions.}
\label{table-ir}
\end{table}

It is evident that {\it pul-km} clearly outperforms the IR-based approaches. In fact, the t-tests indicate statistically significant differences between {\it pul-km} and both {\it ir-i} and {\it ir-p} in all the cases (except in two cases of macro F, one with {\it ir-i} and size 150, and the other with {\it ir-p} and size 25).

\begin{figure}[htb]
\begin{center}
\includegraphics[width=0.65\textwidth]{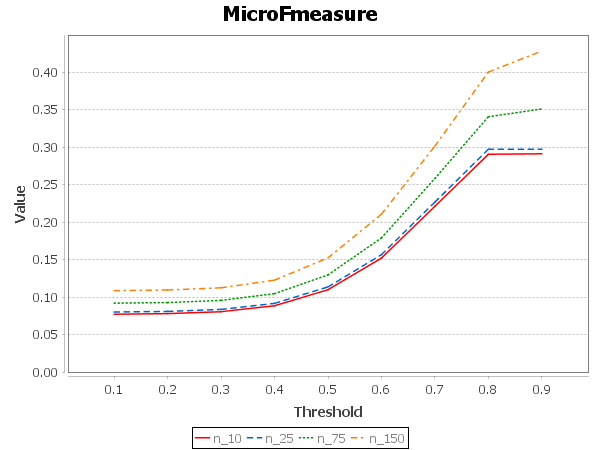}

\vspace{0.1cm}
\includegraphics[width=0.65\textwidth]{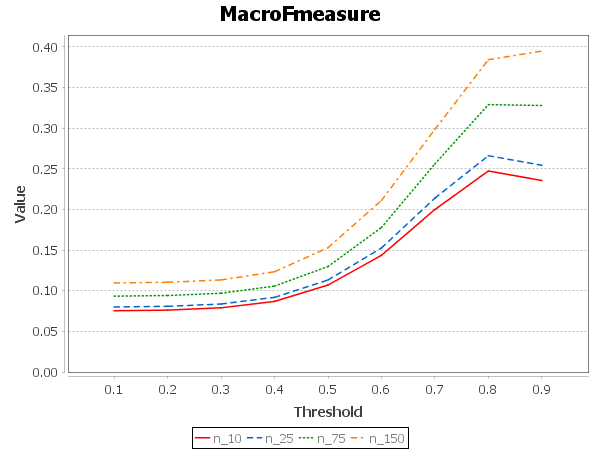}
\caption{Micro and Macro F measures for {\it ir-i} using different thresholds, for a minimum of 10, 25, 75 and 150 interventions.} 
\label{fig-F-IRi}
\end{center}
\end{figure}

Although we are not going to display all the figures showing how {\it ir-i} and {\it ir-p} vary depending on the thresholds being used, we include in Figure \ref{fig-F-IRi} the micro and macro F values for {\it ir-i} (the figures for {\it ir-p} are completely similar) in order to illustrate a clear difference in the behaviour of the IR-based approaches with respect to {\it pul-km}. We can see how the F measures for {\it ir-i} increase as the threshold increases, just the opposite of what it happens with {\it pul-km}. Thus {\it pul-km} works better with low or medium thresholds but {\it ir-i} requires large thresholds. The reason may be in the different interpretation that these thresholds have within each approach, probability of relevance in one case and similarity in the other.

\section{Concluding remarks}
\label{concluding}

In this paper we have proposed an approach to build a system able to recommend/filter documents to the Members of Parliament, which is based on machine learning techniques, more precisely automatic document classification. The source data to train the classifiers are the interventions of the MPs in the parliamentary debates, under the assumption that these interventions reveal information about their political interests and preferences. However, the interventions of an MP only give information about what is relevant for her, but the information about what is not relevant is missing. For that reason our approach uses positive unlabeled learning methods, as we cannot rely on traditional classifiers trained with both positive and negative examples. In this context, we have also proposed a new PUL method, {\it pul-km}, that first obtain a set of reliable negative examples from the set of unlabeled examples (the interventions of the other MPs), and then uses the set of positive and reliable negative examples to train a traditional binary classifier (SVM in our case). Our method to obtain the set of reliable negative examples is based on a modification of the classical K-means algorithm for clustering. We have also considered to complement this procedure with an algorithm to deal with the possible class imbalance problem (using SMOTE for this purpose).

In our experiments, based on a collection of MP's interventions in the Parliament of Andalusia, we have compared {\it pul-km} with other approaches: a baseline approach that considers that all the unlabeled examples are negative examples, another existing PUL method based on Naive Bayes, {\it pul-nb}, and other two methods based on information retrieval that index the collection of interventions and retrieve the MPs which are more similar to the document to be recommended. In all the experiments our approach obtains better results than its opponents, most of the time with statistically significant differences. Therefore, {\it pul-km} appears as a good approach to tackle this recommendation problem. Moreover, the fact that {\it pul-km} clearly outperforms the state-of-the-art {\it pul-nb} is also a strong evidence that it has potential to be useful in other problems where PUL methods are necessary.

Given the results obtained with and without using the SMOTE method to deal with imbalanced data sets, we have observed that its use deteriorates micro F but tends to improve macro F (except in the case of {\it pul-nb}, where macro F is also worsened). This probably means that using SMOTE is only advisable for those MPs having a low number of interventions. Therefore, an interesting future research would be to design strategies to perform ``selective balancing'', i.e. to decide which classifiers (associated to different MPs) would benefit from using methods for balancing data sets. As this operation changes the thresholds that the classifiers need to use to perform better (in our case moving from low thresholds to others located near 0.5, the ``natural'' threshold), another interesting line of research would be to study methods to select different thresholds for different classifiers. Finally, we would also like to explore the use of feature selection methods \cite{maldonado14} (term selection in this case) for our recommendation problem.

\section*{Acknowledgement}

This work has been funded by the Spanish ``Ministerio de Econom\'ia y Competitividad'' under projects TIN2013-42741-P and TIN2016-77902-C3-2-P, and the European Regional Development Fund (ERDF-FEDER).

\end{document}